\documentclass[12pt,a4paper]{article}
\usepackage[latin1]{inputenc}
\usepackage[T1]{fontenc}
\usepackage[english]{babel}
\usepackage{a4,array,latexsym}

\newcommand{\ct}{\cite}

\newcommand{\bc}{\begin{center}}
\newcommand{\ec}{\end{center}}
\newcommand{\bd}{\begin{displaymath}}
\newcommand{\ed}{\end{displaymath}}
\newcommand{\be}{\begin{equation}}
\newcommand{\ee}{\end{equation}}
\newcommand{\ba}{\begin{array}}
\newcommand{\ea}{\end{array}}
\newcommand{\bea}{\begin{eqnarray}}
\newcommand{\eea}{\end{eqnarray}}
\newcommand{\bt}{\begin{tabular}}
\newcommand{\et}{\end{tabular}}

\newcommand{\bp}{\begin{picture}}
\newcommand{\ep}{\end{picture}}
\newcommand{\bfi}{\begin{figure}}
\newcommand{\efi}{\end{figure}}

\begin{document}

\begin{titlepage}

\vspace{1cm}

\centerline{\huge \bf {
Wjj excess from a Higgs boson,}}
\centerline{\huge \bf { 
Composite States}}

\vspace{1cm}

\centerline{\Large\bf  H.B.~Nielsen \footnote{\large\,
hbech@nbi.dk}}

\vspace{1cm}

\centerline{\Large \it The Niels Bohr Institute, Copenhagen,
Denmark}

\vspace{2cm}

\begin{center}{\bf Abstract}\end{center}
\begin{quote}

The newly found \ct{1} dijet peak in the 120 GeV - 160 GeV mass
region produced in co-production with $W$ IS actually a Higgs
boson in spite of the expectation of a different decay pattern for
most Higgses. Our point, however, is that the bound state of $6t +
6\bar{t}$, which we have put forward already in several articles
[4--11], easily could be lighter -- possibly even much lighter --
than half a Higgs mass. Since this bound state, if it existed
should certainly couple strongly to the Higgs, the Higgs would in
this case decay dominantly to two of our bound states. If these
bound states were indeed very light (say $\approx 10$ GeV mass)
their decay products into hadrons would look like two jets, one
for each bound state. That even a very small mass for our bound
state is not unexpected insofar as it is a part of our model that
especially the top-quark-Yukawa coupling is being tuned in so as
to make precisely this bound state of $6t + 6\bar{t}$ become
(approximately) massless. This tuning is a consequence of our
Multiple Point Principle (MPP) [12--14] which states that realized
parameter/couplings values correspond to having a maximally set of
degenerate vacua.

\end{quote}

\end{titlepage}

\clearpage \newpage

\section{Introduction}
Recently there has been seen a three standard deviation ``peak''
\ct{1} in the mass spectrum of dijets in co-production with a
$W$-boson. The mass of around 144 GeV would be perfect for it
being a Higgs boson in the Standard Model, actually agreeing even
with our own prediction for the Higgs mass of 135 GeV \ct{2}, or a
bit lower \ct{3}. However, the decay into $b + \bar{b}$ which is
the decay channel that would show up as a dijet-decay should only
provide 12 fb of the events observed rather than the about $\sim
4$ pb actually making up the possible peak observed. Thus the
Tevatron collaboration \ct{1} asserts that the new particle is not
the Higgs.

Previously we have already discussed in several articles [4--11] a
model in which a cluster of $6t + 6\bar{t}$ is kept together by
Higgs exchange and helped by gluon exchange to form a bound state
with a mass tuned in to be very small, compared say to 12 top
masses. Crudely this bound state should couple as strongly to the
Higgs as 12 top-quarks. Even if this should be a strong
overestimate, it is hard to see that the decay of the Higgs into a
pair of such bound states should not dominate the usually expected
decay modes for the Higgs, at least as long as we compare to
decays involving the rather small bottom-Yukawa-coupling (squared
in the rate) $g_b$, -- as for the $b+\bar{b}$ decay -- or decays
into virtual $Z$ or $W$'s. Now the decay of the Higgs into a
couple of our bound states would -- if these bound states are
sufficiently light and that could easily happen, since they in our
model are {\em finetuned} \footnote{This finetuning comes from our
by now old idea of MPP [12--14] which more recently can be weakly
connected to the idea of a complex action \ct{14,15}. The latter
is in turn forerun by idea more generally of non-locality
[16--18], which we indeed invented trying to find an argument for
MPP. Further C.D.~Froggatt and I established that the existence of
a light bound state is supported by calculations \ct{4,6,7,10}
while our calculation is disputed in Refs.~[19--23]. The idea of
couplings being finetuned was also found in Baby Universe theory
[24-28].} to be light -- look like the Higgs decaying into two
jets. So if our bound state is appreciably lighter than the Higgs
mass, it will cause the Higgs to decay dominantly into a couple of
hadronic jets, i.e. a dijet.

The peak observed by CDF should in our model -- assuming the bound
state to be indeed tuned light -- consist of
\underline{almost\,\,all} the Higgses produced with an associated
event of the type selected, rather than only of the sub-sample
decaying via $b+\bar{b}$ as usually assumed. Very crudely we could
say that the decay into the $b+\bar{b}$ channel is suppressed by a
factor of the order of the bottom-quark-Yukawa coupling $g_b$
squared, $g_b^2$, compared to a channel with coupling of ``order
unity'' as we should expect our two bound states channel to have.
Since the bottom-Yukawa-coupling is of the order of $g_b \approx
1/30$ our decay rate of the Higgs into bound state pairs would be
of the order of $1/g_b^2\approx 900$ times bigger than the decay
rate into $b+ \bar{b}$. Thus an expected into $b+\bar{b}$ decaying
cross-section $\sigma \cdot P(b+\bar{b})\approx 12$ fb would
potentially allow for a cross-section into bound states - which we
suggest is what is observed -- of the order of $900 \cdot 12$ fb
$\approx$ 10 pb. But that would be an over estimate, because in
our model the branching ratio for Higgs into $b + \bar{b}$ would
likely be appreciably diminished compared to the usual picture
(i.e. without our bound state). Even the cross-section for the
peak observed $\approx 4$ pb may be a bit high for Higgs
production at all.

Really the problem is that whole cross-section for Higgs in
co-production with $W$ or $Z$ is more like 0.1 pb than the needed
of the order of 4 pb even if all were observed. In the range of
interesting, Higgs masses say crudely 100 GeV to 200 Gev the
cross-section for co-production with $W$ falls from about 0.25 pb
to 0.02 pb in the usually assumed Standard Model production
mechanism.

With our bound state, or even better other particles in the family
of our $6t + 6\bar{t}$ bound state presented as a virtual particle
in possible mechanisms for Higgs production, a somewhat increased
cross-section for Higgs production compared to the conventional
picture is not excluded. \footnote{In the family of bound states
related to our most important $6t+6\bar{t}$ state we have e.g. a
bound state which is the same one but with one of the top quarks
replaced by a (left) bottom-quark, or we may consider this related
state as a resonance in the scattering of a $W$ and our almost
zero mass state $6t+6\bar{t}$. This resonance we predict to have a
mass in the region 400 GeV to 800 GeV. It could be used to enhance
the effective $WWH$-vertex (i.e. the vertex between two $W$'s and
one Higgs) compared to the Standard Model as usually assumed. This
could come about by the two $W$'s exchanging this resonance in the
400 to 800 GeV range whereby they then become a pair of the light
$6t+6\bar{t}$-bound state virtually. Such a virtual pair couples
very strongly to combine to become a Higgs. In this way we could
obtain a loop diagram involving propagators in the loop
corresponding to our bound states renormalizing the $WWH$-vertex
relevant for the co-production of Higgs ($H$) with $W$.

Even if we say that the full $WWH$-vertex is measured effectively
by the $W$-mass for Higgs four momentum being zero, there is still
the possibility that the suggested loop part of the vertex could
have so much four momentum dependence that a larger coupling at
the four momentum for an on-shell Higgs is possible. Thus such a
correction diagram could at least make the prediction of the rate
of Higgs production with our bound states included more
uncertain.}

With such an increased uncertainty in the up-going direction for
the Higgs cross-section in our model compared to the conventional
estimates the cross-section found for the peak observed of
$\approx$ 4 pb should be comprehensible.

I thus strongly suggest that what has been seen by CDF IS the Higgs,
where the Higgs dominantly decays into two very light
bound states looking like jets and therefore does not match with usual
theoretical expectations for a Higgs.

In the following section we discuss a couple of arguments against
that the observed peak should be due to a Standard Model Higgs. In
the last section I shall conclude and resume stressing that our
model with the bound state was indeed developed quite without any
indication of the Higgs decaying into hadrons rather than in the
conventional way with $W$ etc. except perhaps that there were some
indications that the Higgs seen by LEP were effectively broad
\ct{29,30}.

\section{Discussion of arguments against the peak being a Higgs}

The Tevatron group investigated whether the dijets in the peak
were more rich in b-jets than the surrounding mass regions as
would have been the case, if the peak were due to $b+\bar{b}$
decay (of a conventional Higgs). There were NO significant excess
of b-jets. If the jets are really decay products of our $6 t +
6\bar{t}$ bound states, the question of finding an excess of
b-jets -- as was said not to have been found -- would depend on
whether the decay of our bound state would go preferentially to
b-quarks. In the approximation of strong interactions dominating
so that only gluons cause the annihilation of the several pairs of
$t\bar{t}$ in the bound state there would be no special b-excess.
So in this first approximation -- gluon dominance -- our model
agrees with the lack of b-excess in the peak. If, however,
$W$-exchanges should play a role in the disappearance of the top
and anti-top quarks in the bound states, an excess of bottom
quarks in the decay is expected because of the dominant weak ($W$)
transition from top to bottom.

The genuine $b+ \bar{b}$ decay which also looks like a dijet would
as we crudely estimated be down by a factor 900 and would hardly
be observable in data.

Another argument against that the data with the peak revealing the
Higgs is that a neural network program has already looked for
Higgses and generally careful Higgs studies have also been done.
However, neural networks looking for Higgses will likely look
mainly for decays that are with the characteristic lepton and
missing transverse energy. Thus such programs trained to look for
the conventional Higgses will presumably not accept the dijet
decaying Higgs as a very promising candidate. Actually for the
conventional Higgs only the channel $b+\bar{b}$ with relatively
small branching ratio, with lots of background will be of the
dijet type. If there are no signs of b-quarks the neural network
will presumably rank the chance that a dijet decay be a Higgs as
very small.

The conventional Higgs decay modes will presumably be strongly reduced
by the Higgses decaying into pairs of our bound states. So we predict that
the conventional  Higgs searches will have only little success.

\section{Conclusion:  Higgs  found by CDF  decay mainly into
two jet-like bound states! }

In the present article I have pointed to that provided our bound
state of $6 t+6\bar{t}$ [4--11] is very light in accordance with
our finetuning principle, ``multiple point principle''\,(MPP)
[12--14], it is no longer a problem that the newly found particle
co-produced with $W$ could be a Higgs! Indeed our point is that
the very light -- exceptionally finetuned to be light -- bound
state of $6+\bar{t}$ couples strongly to the Higgs. Therefore the
Higgs decays dominantly into a pair of such bound states and
likely shows up as decaying into two jets. Thereby usually
believed Higgs decays will not be able to compete (in branching
ratio) and a conventional Higgs search might lead to very few
results. Of course, also a broader Higgs width is expected in our
model. Possibly the width seen in the observed peak extending from
around 120 Gev to 160 GeV could easily in our model reflect the
genuine decay rate rather than uncertainties in the mass
measurement.

It should be stressed that the model with a six top plus six
anti-top bound state propagated in the present article were NOT
made up for the purpose of the CDF-dijet-mass-peak. Rather we have
speculated and published our bound state idea as an extension of
our Multiple Point Principle (MPP) originally put in Refs.~
[12--14] as a fundamental law stating that Nature seeks out values
of parameters/couplings that correspond to having maximally
degenerate vacua. As the transition between the phases
corresponding to the various vacua are first order, our MPP also
embodies a finetuning mechanism (due to having finite heats of
melting and fusion if one thinks in terms of the somewhat
analogous phase diagram for water).

Subsequently MPP has been used to explain the low weak scale
\ct{6,31}, or originally to fit gauge coupling constants [12--14],
and  to provide a model for dark matter \ct{11}, and to understand
the value of the cosmological constant \ct{32,33}. Putting into
our already published model with the 6 top + 6 anti-top bound
state the option that the mass of it is lower than half the Higgs
mass leads immediately to the purely hadronic and likely dijet
decay of the otherwise Standard Model Higgs. It should be stressed
that our model is -- or at least could be -- purely Standard Model
in the sense that no new fields in addition to those of the
Standard Model go into our model. The only deviation is not truly
a deviation, but rather the addition of extra information:

We postulate that the coupling constants in the Standard Model are
tuned to obey relations between them guaranteeing that not only
one, but rather three different vacua shall have their
vacuum-energy densities (or cosmological constants) be very small
(essentially zero).

 One of these three vacua is supposed -- in our model [4--11] --
to be one deviating from another in which we live by having a
Boson condensate of our above discussed bound state. It is the
tuning of e.g. the top-Yukawa-coupling to make the vacua with and
without this condensate energetically degenerate that is supposed
also to lead to the bound state mass being small.

Contrary to the general belief that it is not possible to
understand dark matter without extending the Standard Model, our
model can incorporate dark matter in the form of pea-sized white
dwarf stars -- or of similarly strongly compressed matter -- in
which the vacuum {\em with} the Bose-condensate of the very
particle coming out of the main Higgs decay in the prevailing
vacuum inside the sea.

A potential further project might be to fit the CDF-peak discussed
together with the LEP findings \ct{29,30} as a tail of now assumed
broad Higgs.

A possible way to confirm our model would be in the CDF-data to
look for whether the jets in the dijet should have a mass peak,
i.e. for the jets for which total mass of the jet of hadrons can
be calculated sufficiently accurately over an abundance of jets
with a certain mass (of course, the mass of our NBS bound state,
which we estimate to be extremely narrow).

\section{Acknowledgement}
It is pleasure to thank Ivan Andric (Rudjer Boskovic Institute)
for the first information about the discussed CDF-peak and for
further discussions during my visit to Zagreb. I thank Larisa
Jonke for funds. Also I thank my collaborators on the work about
the bound state from which almost all the present article could
have been extracted, in particular, Roman Nevzorov (Hawaii Univ.)
for telephone conversations about the Higgs decaying into our
bound state at a very early stage in our developments of the bound
state theory. I am thankful of Don Bennett, C.R.~Das and Larisa
Laperashvili for some corrections during the proof-reading. Also I
want to thank Li Shi-Yuan (School of Physicis, Shandong
University) and P.R.~China for the collaboration long ago in ca
2000 on the question of the decay of our $6t + 6\bar{t}$ bound
state.

\end{document}